\documentclass[aps,prl,reprint,superscriptaddress,amssymb,amsmath]{revtex4-2}
\usepackage{graphicx}
\usepackage{dcolumn}
\usepackage{bm}
\usepackage{cancel}
\usepackage{enumitem}
\usepackage[T1]{fontenc}
\usepackage{lmodern}

\begin{document}


\title{Exploiting the Planck-Einstein Relation}


\author{R. Engels}
\email{r.w.engels@fz-juelich.de}
\affiliation{Institut f\"ur Kernphysik, Forschungszentrum J\"ulich, J\"ulich, Germany}
\author{M. B\"uscher}
\affiliation{Peter-Gr\"unberg Institut, Forschungszentrum J\"ulich, J\"ulich, Germany}
\affiliation{Institut f\"ur Laser- und Plasmaphysik, Heinrich-Heine-Universit\"at D\"usseldorf, D\"usseldorf, Germany}
\author{P. Buske}
\affiliation{Institut f\"ur Kernphysik, Forschungszentrum J\"ulich, J\"ulich, Germany}
\affiliation{now at: Lehrstuhl f\"ur Technologie Optischer Systeme, RWTH Aachen University, Aachen, Germany}
\author{Y. Gan}
\affiliation{Institut f\"ur Kernphysik, Forschungszentrum J\"ulich, J\"ulich, Germany}
\affiliation{Fachhochschule Aachen, Campus J\"ulich, J\"ulich, Germany}
\author{K. Grigoryev}
\affiliation{Institut f\"ur Kernphysik, Forschungszentrum J\"ulich, J\"ulich, Germany}
\author{C. Hanhart}
\affiliation{Institut f\"ur Kernphysik, Forschungszentrum J\"ulich, J\"ulich, Germany}
\affiliation{Institute for Advanced Simulation, Forschungszentrum J\"ulich, J\"ulich, Germany}
\author{L.~Huxold}
\affiliation{Institut f\"ur Laser- und Plasmaphysik, Heinrich-Heine-Universit\"at D\"usseldorf, D\"usseldorf, Germany}
\author{C. S. Kannis}
\affiliation{Institut f\"ur Kernphysik, Forschungszentrum J\"ulich, J\"ulich, Germany}
\affiliation{III. Physikalisches Institut B, RWTH Aachen University, Aachen, Germany}
\affiliation{JARA-Fame, Forschungszentrum J\"ulich and RWTH Aachen University, Germany}
\author{A. Lehrach}
\affiliation{Institut f\"ur Kernphysik, Forschungszentrum J\"ulich, J\"ulich, Germany}
\affiliation{III. Physikalisches Institut B, RWTH Aachen University, Aachen, Germany}
\affiliation{JARA-Fame, Forschungszentrum J\"ulich and RWTH Aachen University, Germany}
\author{H. Soltner}
\affiliation{Zentralinstitut f\"ur Engineering, Elektronik und Analytik, Forschungszentrum J\"ulich, J\"ulich, Germany}
\author{V. Verhoeven}
\affiliation{Institut f\"ur Kernphysik, Forschungszentrum J\"ulich, J\"ulich, Germany}

\date{\today}

\begin{abstract}
The origin of quantum physics was the discovery of the base unit of electromagnetic action $h$ by Max Planck in 1900 when he analyzed the experimental results of the black body radiation. This permitted Albert Einstein a few years later to explain the photoelectric effect by the absorption of photons with an energy of $E = h \cdot f$. We exploit the Planck-Einstein relation in a new type of fundamental spectroscopic measurements of direct transitions between two states with energy differences of about 10 neV and induced frequencies of a few MHz. Employing a Lamb-shift polarimeter and a Sona transition unit, featuring a relatively simple magnetic field configuration of two opposing solenoidal coils, we were able to determine $f$ and measure $E$ independently. Only resonances corresponding to integer multiples of Planck's constant $h$ were observed in our setup, which can very well be explained quantitatively by the Schr\"odinger equation. This new method beautifully demonstrates the quantization in the micro-cosmos and allows one to measure the hyperfine splitting energies between the substates with $F=1$ and $m_F = -1, 0, +1$ of metastable hydrogen atoms as function of a magnetic field and, thus, to investigate the influence of QED corrections on the Breit-Rabi diagram.
\end{abstract}


\maketitle

\section{Introduction}
When Max Planck described the black body radiation in 1900~\cite{planck}, he saw himself forced to introduce a new constant that is nowadays named in his honor, the Planck constant $h$. He suggested that an electrically charged oscillator can change its energy only in a discrete portion $E$ that is proportional to its frequency $f$, thus
$$E = h \cdot f.$$
Accordingly, $h$ is the smallest quantity of action. This was the starting point for quantum physics and directed Albert Einstein in 1905 to the description of the photoelectric effect~\cite{einstein} in terms of the absorption of light quanta, later called photons. Since 2019 the Planck constant is defined~\cite{cgpm} as
$$h = 6.62607015 \cdot 10^{-34}\,\, {\rm Js} = 4.135667696 \cdot 10^{-15}\,\, {\rm eV s}.$$
It is one of the seven fundamental constants employed to define the base units of the SI system. A few years after Einstein's famous work, Niels Bohr~\cite{bohr} showed that the absorption of a photon of a suitable energy by an atom induces its transition into an excited state and that the decay of this excited state back into the ground state leads to the emission of a photon of the same energy. The energy difference between these states, $\Delta E$, is again $h \cdot f$.\\

Here we report on an experiment to determine the oscillation frequency of the magnetic field ${f {\sim}}$MHz and to measure the energy difference between single quantum states at ${\Delta E {\sim} 10^{-8}}$ eV independently. These tiny energy differences can be found in the Breit-Rabi diagram~\cite{breit} (see Fig.~1) between hyperfine substates of metastable hydrogen atoms in the 2S\textsubscript{1/2} state in the Zeeman region~\cite{zeeman}, which is realized for magnetic fields below $$B_c = \frac{\Delta E_{HFS}}{2 \mu _B} = 6.34 \,\, {\rm mT,}$$
where $\Delta E_{HFS} = 7.34 \cdot 10^{-7}$ eV denotes the hyperfine splitting energy of the 2S\textsubscript{1/2} state and $\mu_B = \frac{e \hbar}{2 m_e} = 9.274 \cdot 10^{-24}$ J/T is the Bohr magneton. 

\begin{figure}
	\centering
	\includegraphics*[width=0.47\textwidth]{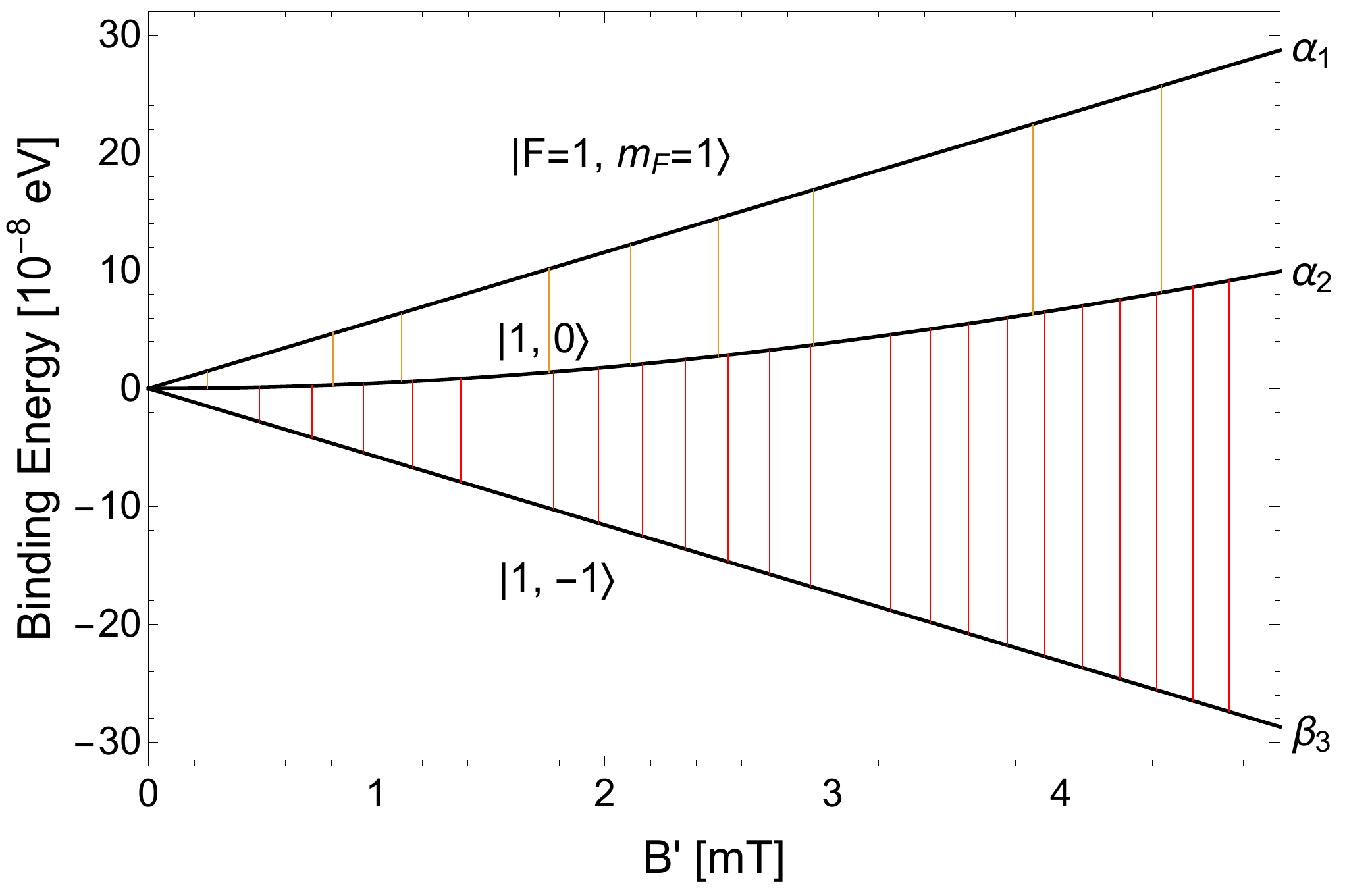}
	\caption{ The Breit-Rabi~\cite{breit} diagram shows the relative binding energies of hyperfine substates as function of an external magnetic field $B'$. At low magnetic fields in the Zeeman region~\cite{zeeman} of the metastable 2S\textsubscript{1/2} state the energy differences between the substates with $F = 1$ and the projection $m_F = +1, 0, -1$ along an external field $B'$ are in the order of $10^{-8}$ up to $10^{-7}$ eV. If a radio-frequency $f = 3.54$ MHz corresponding to an energy of $1.464 \cdot 10^{-8}$ eV is induced, then (multi-)photon-induced transitions ($\alpha_2 \leftrightarrow \alpha_1$: yellow lines, $\beta_3 \leftrightarrow \alpha_2$: red lines) are only possible at magnetic fields corresponding to energy differences between these substates of $\Delta E = n \cdot h \cdot f$.\label{fig1}}
\end{figure}

In this case, the total electron spin $J=1/2$ of the 2S state consists only of the electron spin. It couples with the nuclear spin $I=1/2$ of the proton to the total angular momentum $F=J+I$ of the hydrogen atom, which can thus be either 1 or 0. For $F=1$ three substates exist with the projection $m_F = +1, 0, -1$ onto the external magnetic field direction as quantization axis named $\alpha_1$, $\alpha_2$ and $\beta_3$, respectively. The $F = 0$, $m_F = 0$ substate is called $\beta_4$. Starting with a beam of atoms in a single hyperfine substate, their interaction with photons changing the occupation numbers was studied in detail in the experiment outlined below.

\section{Experimental Setup and Results}
\begin{figure*}
	\centering
	\includegraphics*[width=\textwidth]{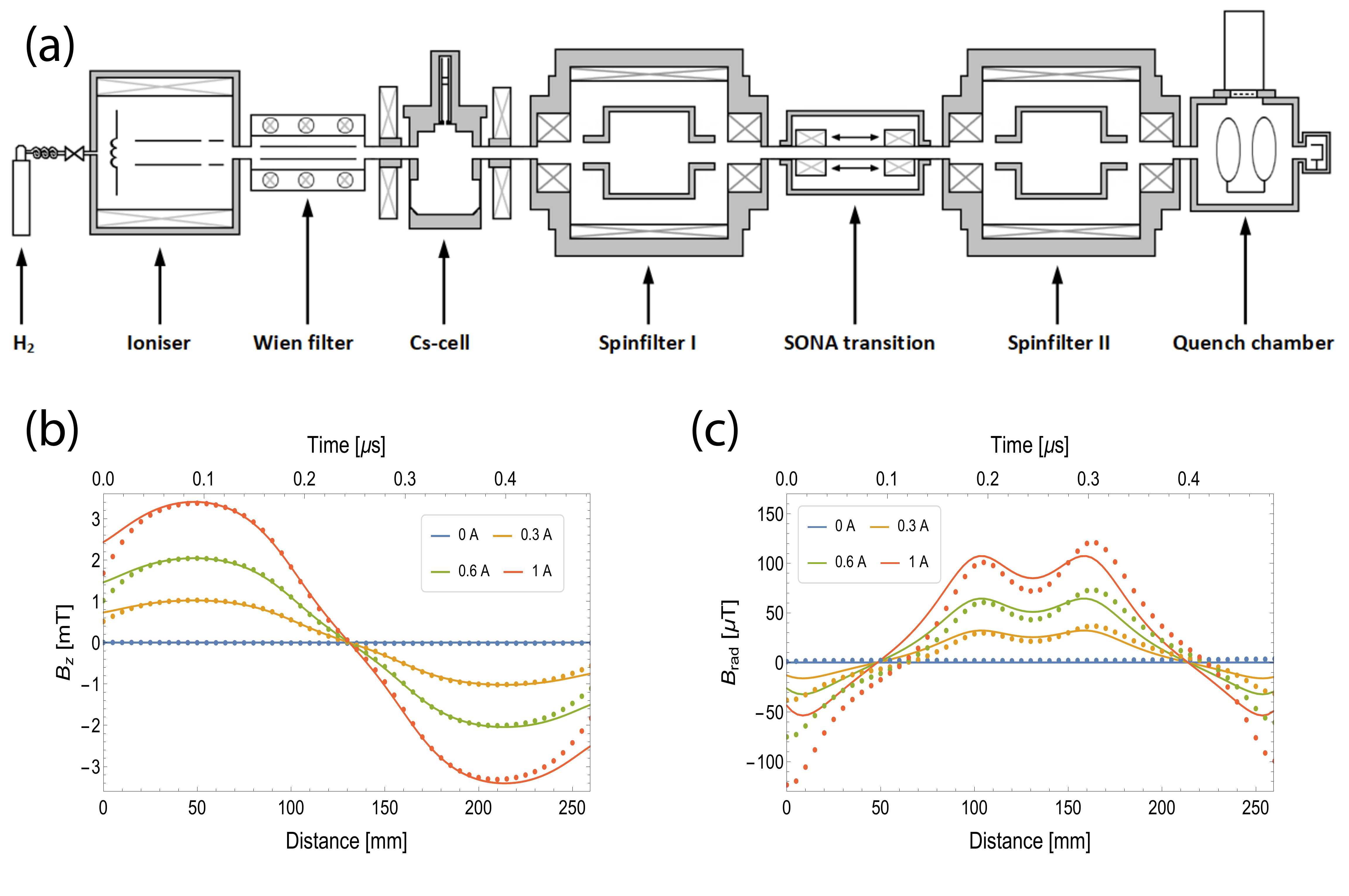}
	\caption{The experimental setup (a) and the calculated (line) and measured (dots) longitudinal (b) and radial (c) magnetic fields at a distance of $\rho = 3$ mm from the beam axis $z$ for different currents in the coils of the Sona transition unit and a distance of 60 mm between the coils, which corresponds to a distance between the center of the Sona coils of 160 mm. The calibration of the maximum magnetic field $B_{max}$ in the center of the Sona coils on the current $I$ is $B_{max}(I) =  3.37\ \rm{mT/A} \cdot I + 0.011\ \rm {mT}$.  \label{fig2}}
\end{figure*}

The experiment is based on the components of a Lamb-shift polarimeter that is frequently used to measure the polarization of hydrogen and deuterium atoms~\cite{engels}, molecules~\cite{engels3} and their ions. 
In a first step molecular hydrogen is ionized in an electron-impact ionizer, and the resulting protons are accelerated to kinetic energies $E_p$ between 1 and 2 keV (see Fig.~\ref{fig2}). A Wien filter with its crossed electric and magnetic fields is used to filter out ${\rm H_2 ^+}$ ions from protons due to their different velocities before they reach the cesium cell. Here, by charge exchange with cesium vapor that is obtained by heating a small amount of cesium at the bottom of the cell, metastable hydrogen atoms in all four hyperfine substates are produced. After that, a spinfilter quenches all metastable atoms into the ground state and allows for special resonance conditions only metastable atoms in the states $\alpha_1$ or $\alpha_2$ to be transmitted~\cite{mckibben}. The next component is a ``Sona transition unit''~\cite{sona} that consists of two 10 cm long solenoids providing opposing longitudinal magnetic fields with a zero crossing in the gap between the coils. If hydrogen atoms in the state $\alpha_1$, i.e.~electron and proton spin are parallel to the external magnetic field along the beam axis, will leave the spinfilter, then it is possible to invert the external field direction faster than the Larmor precession of the total spin $F$. In this non-adiabatic case, the spins keep their direction along the beam axis $z$ when the magnetic field is inverted. For this to happen the magnetic field gradient around the zero-crossing needs to fulfill the requirement~\cite{sona} 
$$\frac{dB_z}{dz} < \frac{8\ v_H\ m_e}{e\ r^2},$$ 
where $v_H \approx\sqrt{2 E_p /m_p}$ denotes the velocity of the hydrogen atoms, $m_p = 1.672 \cdot 10^{-27}$ kg the  proton mass, $m_e = 9.109 \cdot 10^{-31}$ kg and $e = 1.602 \cdot 10^{-19}$ C are the electron mass and the unit charge, respectively, and $r$ the beam radius. This requirement on the field gradient can be realized by a proper choice of the magnetic field $B$, the beam energy $E_p$ and the distance between the Sona coils. For example, a kinetic energy of the protons of 1.28 keV corresponds to a velocity of the hydrogen atoms of $v_H = 4.93 \cdot 10^{5}$ m/s. For a given beam radius of $r = 1$~cm the gradient must be below $dBz/dz < 2.2$ mT/cm, which corresponds to a current of 5~A in the coils, if their distance is 60 mm.

With such a setting all atoms in state $\alpha_1$ are transferred into state $\beta_3$ and vice versa, i.e.~the projection of $F$ onto the quantization axis is changed from $m_F = +1$ into $-1$. This very efficient method of changing occupation numbers was used in polarized ion sources of the Lamb-shift type~\cite{bechtold} and is still in use at optically pumped ion sources~\cite{zalenski}.

Behind the Sona transition unit another spinfilter is used to analyze the amount of metastable atoms in the single $\alpha$ substates. For example, if the first spinfilter is set to transmit only metastable atoms in the substate $\alpha_1$, the Sona transition unit transfers these atoms into the substate 
$\beta_3$ and none of them is transmitted through the second spinfilter (see Fig.~\ref{fig2}). If the Sona settings transfer some atoms back into the $\alpha_1$ or $\alpha_2$ substate, they pass the second spinfilter until the corresponding resonant conditions allow them to pass through. Afterwards, these metastable atoms reach the quenching chamber where they are quenched into the ground state by the Stark effect due to a strong electric field~\cite{stark}. The amount of produced Lyman-$\alpha$ photons is registered with a photomultiplier as function of the current through the Sona coils. In Figs.~3 and 4 (blue graph) it is demonstrated that for some magnetic flux density values metastable atoms in substate $\alpha_1$ or $\alpha_2$ appear in the second spinfilter in contrary to naive expectations~\cite{sona}. 
\begin{figure}
	\centering
	\includegraphics*[width=0.49\textwidth]{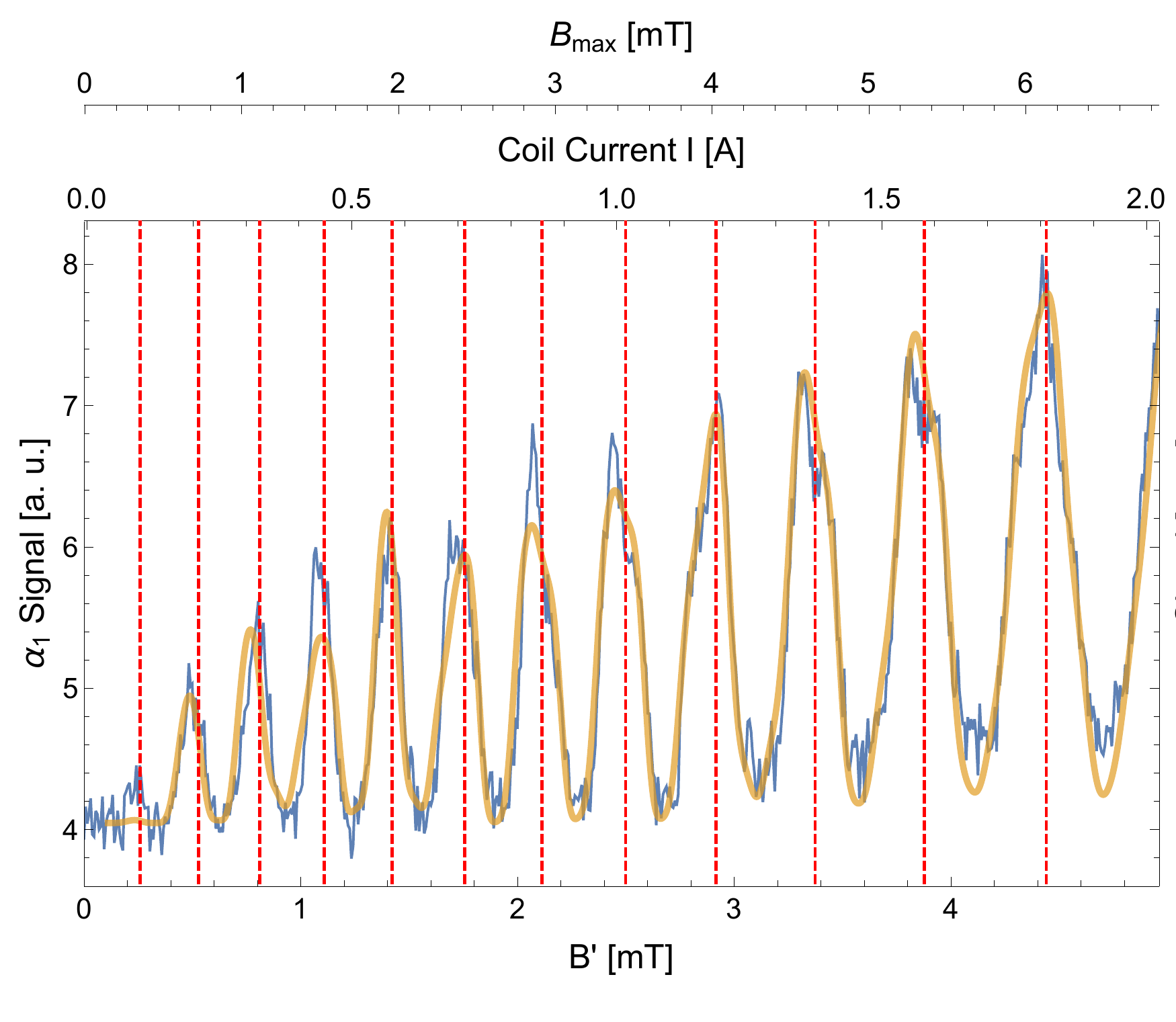}
	\caption{The measured relative amount of metastable hydrogen atoms in the hyper-fine substate $\alpha_1$ behind the Sona transition unit (blue) as function of the the effective magnetic field $B'$ up to 5 mT, the current in the Sona coils and the corresponding max.~magnetic field $B_{max}$ in the center of the Sona coils. The offsets between magnetic fields and the current stem from unshielded magnetic stray fields in the interaction region. In addition, the prediction of the peak centers due to the Breit-Rabi diagram (red dashed lines) and the solution of the time-dependent Schr\"odinger equation (yellow) for the measured magnetic field distribution B(z) and the velocity $v_H=4.93 \cdot 10^5$ m/s ($E_{beam} = 1.28$ keV) are presented. \label{fig3}}
\end{figure}

\begin{figure}
	\centering
	\includegraphics*[width=0.49\textwidth]{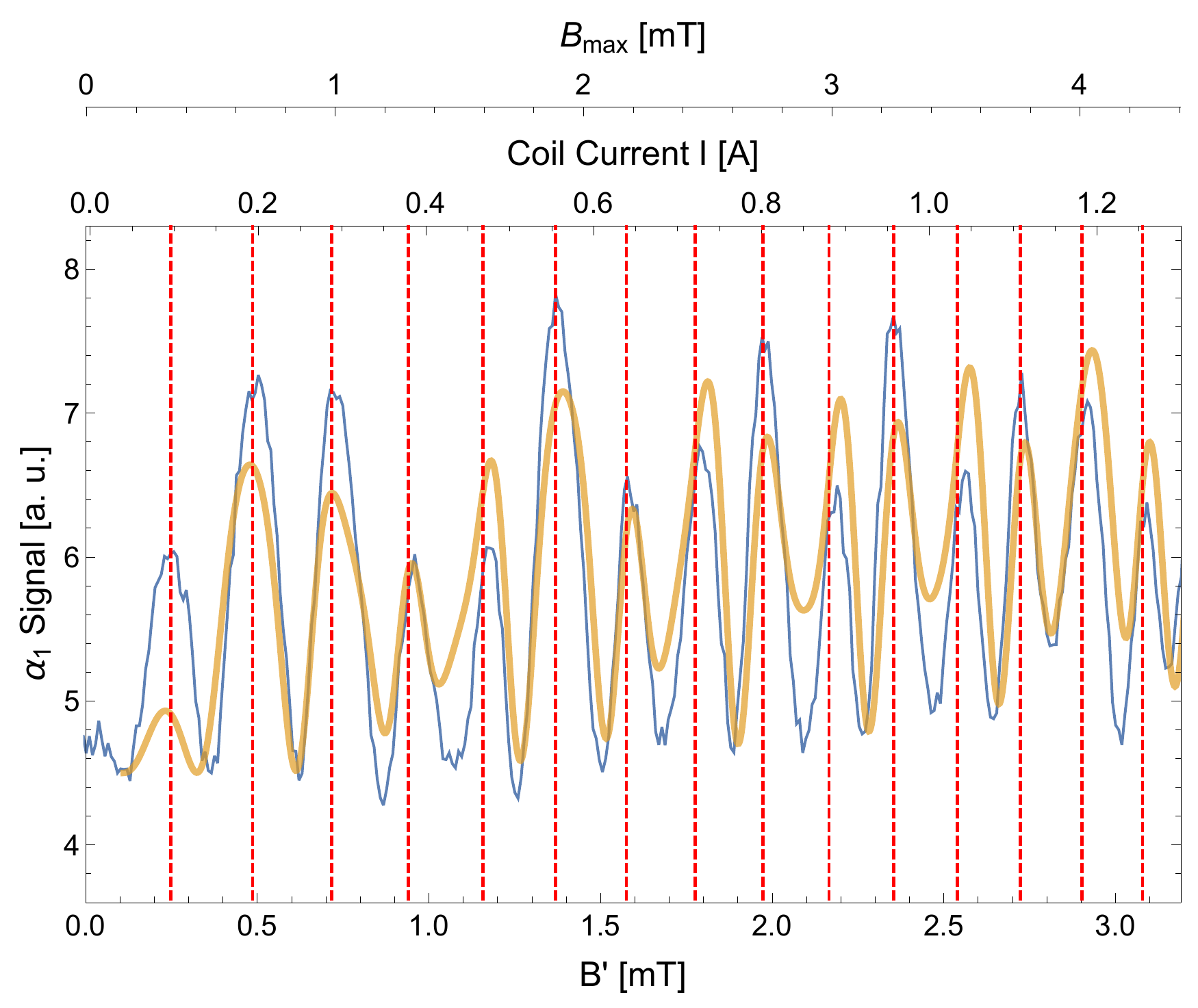}
	\caption{The measured relative amount of metastable hydrogen atoms in the hyperfine substate $\alpha_2$ behind the Sona transition unit (blue) as function of the effective magnetic field $B'$ up to 3.2~mT, the current in the Sona coils and the cor\-res\-ponding max.~magnetic field $B_{max}$ in the center of the Sona coils. Other details are found in the caption of Fig.~3.}
\end{figure}

We note that a similar effect has been reported for a beam of hydrogen atoms in both $\alpha$ substates of the metastable 2S\textsubscript{1/2} state ~\cite{hight,robiscoe,garisto} and even of the 1S\textsubscript{1/2} ground state~\cite{kponou,zalenski2} before, but the occupation numbers of single substates have not been investigated in detail until now.

\section{Theoretical description and discussion}
Since the individual atoms move on a straight line with a constant velocity through the apparatus, we can equivalently describe the system as an atom at rest, experiencing a magnetic field that varies in time. The corresponding Hamilton operator describing the hyperfine substates of a hydrogen atom in an external magnetic field $B$ is~\cite{antishev,milstein}

\begin{eqnarray*}
\begin{split}
H(t) &= \Delta E_{HFS} \bm{I} \cdot \bm{J} - \bm{\mu_{atom}} \cdot \bm{B(t)} \\
&= \Delta E_{HFS} \bm{I} \cdot \bm{J} - (g_J \mu_B \bm{J} + g_p \mu_K \bm{I}) \cdot \bm{B(t)}\\
&= H_0 + V(t),
\end{split}
\end{eqnarray*}
where $H_0$ denotes the first, time-independent term. Here $g_J = - 2.001$ and $g_p = 5.586$ denote the g-factor of the electron and the proton, respectively, and $\mu_K = \frac{e h}{4 \pi m_p} = 1.41 \cdot 10^{-26}$ J/T is the nuclear magneton.

The only tuneable quantity is the magnetic field $\bm{B(t)}$. Its time dependence can be replaced by $B(z)$, i.e. the dependence of the magnetic field along the beam direction $z$ due to the tuneable velocity of the hydrogen atoms (see Fig.~\ref{fig2}). The radial component $B_{rad}$ of the flux density at a radius $\rho$ can be calculated from the gradient of the longitudinal field $B_z$ via $$ B_{rad}(z,\rho) = -\frac{dB_z}{dz}  \cdot \frac{\rho}{2}$$ and, of course, directly measured.

Note that only $B_{rad}$ can induce the observed transitions, since both terms $J_z \cdot B_z$ and $I_z \cdot B_z$ are diagonal in the basis used. Using the time-dependent Schr\"odinger equation and the measured $B_{rad} (\rho, z)$ with the Hamiltonian given above and expanding the solution in terms of the unperturbed wave functions (found from solving the time-independent Schr\"odinger equation with $V(t)=0$),
$$ \Psi (x,t) = \sum_{n=1} ^4 c_n (t) e^{-i E_n t/\hbar} |n\rangle  \quad {\rm with} \quad H_0 |n\rangle = E_n |n\rangle$$ one finds for the expansion parameters the equation $$i\hbar \frac{dc_{k} (t)}{dt} = \sum_{n=1} ^4 c_n (t) e^{\frac{-i (E_n -E_k) t}{\hbar}} \langle k | V(t) |n\rangle,$$ where the sum runs over the four basis states of the $2S$ level. The only approximation applied to derive this differential equation is that all states not belonging to the $2S$ state are neglected, which is well justified given the large energy differences. For the discussion below it is important to note, that the matrix element $\langle k | V(t) | n \rangle$ can only yield a contribution for quantum numbers $m_F$ differing by $\pm 1$.
 
This equation may be solved by discretizing in the time variable for the known magnetic field $B_{rad} (\rho,z)$. To find the relative occupation numbers for the different substates, one now only needs to integrate the resulting wave functions over the beam profile, here assumed to be of Gaussian form with a spread $\sigma = 0.5$ cm. In Figs.~\ref{fig3} and 4 the results for $\alpha_1$ and $\alpha_2$ are compared to the experimental data. The radial beam density profile does not affect the peak positions at the corresponding magnetic fields, but influences the amplitude and the shape like it can be especially observed in Fig.~4. Even if the relative velocity distribution of the metastable atoms in the beam is below $\pm 2\%$, only an increased resonance width but no peak deformations appears.

It is intriguing that the locations of the peaks shown in Figs.~\ref{fig3} and 4 can be straightforwardly understood in a particle picture. It is the general understanding that during their passage from one Sona solenoid to the other all atoms are transferred from state $\alpha_1$ into $\beta_3$. However, in addition they can absorb a photon to be excited into state $\alpha_2$ and a second photon to reach state $\alpha_1$ again. These photons are induced due to an oscillation of the radial magnetic field seen by the metastable hydrogen atom in its center-of-mass system during the flight from one Sona coil to the other. Those transitions are by far most efficient, when $\Delta E = n \cdot h \cdot f$ holds. Therefore, every time the total magnetic field inside the solenoids corresponds to an energy difference between the states $\alpha_1$ and $\alpha_2$ that multiplied with $1/f$ gives an integer multiple of $h$ (see Fig.~\ref{fig1}), this transition can appear as a peak in a scan like the one shown in Fig.~\ref{fig3}, i.e.~this transition is induced by multi-photon transitions. The transition from $\alpha_2$ into $\alpha_1$ is only possible, if the state $\alpha_2$ has been populated due to a $\beta_3 \leftrightarrow \alpha_2$ transition before, for otherwise the transition matrix element vanishes. Due to the fact that the energy differences between the states $\beta_3 \leftrightarrow \alpha_2$ and $\alpha_2 \leftrightarrow \alpha_1$ are not equal for the same magnetic field, the resonance shapes are a convolution of both transitions. The corresponding deformation of the single resonances are obvious in Fig.~\ref{fig3} for the measured and the calculated data.
When the second spinfilter allows only metastable hydrogen atoms in the substate $\alpha_2$ to be transmitted, the transitions from $\beta_3 \rightarrow \alpha_2$ will dominate the occupation numbers of the substate $\alpha_2$ and the transitions $\alpha_2 \rightarrow \alpha_1$ are responsible for an additional deformation of the resonances (see Fig.~4). 

The energy of the exchanged photons can be determined with different methods:\\
As described before, the radial magnetic field $B_{rad} (z,\rho)$ can be transformed with the known beam velocity $v$ into $B(t,\rho)$. Thus, a Fourier analysis of the radial radio-frequency seen by the atom during its time-of-flight $\Delta t$ through the Sona coils yields for a beam velocity of $v_H = 4.93 \cdot 10^5$ m/s a first harmonic frequency of $f_0 = 1.76$ MHz and a second harmonic at $f = 3.52$~MHz. Only the second harmonic corresponds to a full oscillation between the Sona coils, i.e.~a photon, and can induce radio-frequency transitions between the hyperfine substates. All harmonic frequencies depend only on $v_H$ and the geometry of the coils and are independent of an increasing longitudinal magnetic field strength that only changes the amplitude of the oscillation, i.e.~the number of photons. The corresponding wavelength $\lambda$ of the second harmonic frequency $f$ can be calculated due to the equation $$v_H = \lambda \cdot f$$ and is $\lambda = 140$ mm in the example of Fig.~\ref{fig2}. \\
Thus, the static radial magnetic field component in the laboratory system induces an oscillating radial magnetic field in the center-of-mass system of the atom and the frequency $f=1/\Delta t $  corresponds for a given magnetic field geometry that defines $\lambda$ only on the velocity $v_H$ of the beam. This allows one to control the induced frequency in a wide range and to observe direct transitions between quantum states with an energy difference of a few MHz for $v_H \sim 10^5$~m/s down to 10~kHz for $v_H\sim10^3$ m/s. In addition, the distance between the Sona coils can be increased to decrease the frequency even further.

The non-linearity of the binding energies of the hyperfine substates $\alpha_2$ as function of the magnetic field permits a more precise calibration method. With an effective magnetic field $B'$ as input, the Breit-Rabi formula including QED corrections~\cite{antishev} can be employed to calculate the energy difference between the hyperfine substates (see Fig.~\ref{fig1}). For the measurement shown in Figs.~\ref{fig3} and 4 the first peak appears at a current of 0.09~A through the Sona coils that corresponds to a maximum magnetic field of $B_{max} = 0.31$ mT inside the coils. For the next resonances the differences between the corresponding magnetic fields increase for the $\alpha_2 \leftrightarrow \alpha_1$ transitions with larger $n$ and decrease for $\beta_3 \leftrightarrow \alpha_2$. At the same time the energy differences between the states $\Delta E = \Delta E_{(B')} = n \cdot h \cdot f$ depend only on the number of the peaks $n$ and the induced radio-frequency $f$, which is constant for a constant velocity of the atoms and a fixed experimental setup. In this way, a calibration for the effective magnetic field $B' =  0.723 \cdot B_{max} + 0.004 $ mT could be found with an offset at zero current due to not perfectly shielded external fields. For a scan from $I=0$ up to 2 A in the Sona coils $n=12$ resonances are found for the transitions between $\alpha_2 \leftrightarrow \alpha_1$ (Fig.~\ref{fig1} and~\ref{fig3}) and $n=26$ for $\beta_3 \leftrightarrow \alpha_2$. In this way, $\Delta E_{(n=1)} = (1.462 \pm 0.005)\cdot 10^{-8}$ eV or $f= 3.536 \pm 0.012$ MHz was determined with the current setup, which corresponds to an absolute uncertainty of $\Delta E {\sim} 10^{-11}$ eV or $\Delta f {\sim} 10$ kHz.

Further improvements, e.g.~higher proton beam intensities for better statistics, more convenient setups of the magnetic fields in the Sona transition unit~\cite{zalenski2} and better shielding of the external fields, should permit a reduction of the uncertainty by another factor 10. If the beam energy and, thus, the velocity of the protons is changed in the ionizer, $\Delta E$ and $f$ are modified too, but the fraction remains to be the Planck's constant $h$. Thus, the statistical uncertainty can be decreased further to measure the hyperfine splitting energies of these states as function of a (small) magnetic field down to $10^{-13}$ eV or 100 Hz, correspondingly.  

The Schr\"odinger equation itself has the potential to deliver an analysis that can be even more precise: If the occupation numbers of the hyperfine substates are measured like in Fig.~\ref{fig3}, it is even possible to calculate the longitudinal magnetic field distribution $B(t)$ along the beam axis in the Sona coils from these data inversely. A Fourier expansion of the deduced radial magnetic field delivers directly the frequencies $f$ of the oscillation of the radial component $B_{rad}$, and, therefore, the energy difference $n \cdot \Delta E$ between the substates for the single resonances and the corresponding magnetic fields $B'$.
 
The same method works for metastable deuterium atoms as well. Here, the Sona transition unit transfers all atoms in the state $\alpha_1 (F=3/2, m_F = +3/2)$ into $\beta_4 (m_F =-3/2)$ and vice versa; the spinfilter is able to separate the $\alpha$ substates $\alpha_1$, $\alpha_2 (m_F = +1/2)$, and $\alpha_3 (m_F = -1/2)$.  

The Breit-Rabi formula shows that for protons (deuterons) the sum of the hyperfine splitting energies between the substates $\alpha_2 \leftrightarrow \alpha_1$ and $\beta_3$~$\leftrightarrow$~$\alpha_2$ ($\alpha_2 \leftrightarrow \alpha_1$, $\beta_3 \leftrightarrow \alpha_2$, and $\beta_4 \leftrightarrow \alpha_3$) as function of the magnetic field $B'$ is a linear function, i.e.~$E_{(\alpha_2 \leftrightarrow \alpha_1)} (B') + E_{\beta_3 \leftrightarrow \alpha_2} (B') = 2 (-g_J \mu_B +g_p \mu_K) B'$~(Deuterons:~$E_{(\alpha_2 \leftrightarrow \alpha_1)}(B')+E_{(\alpha_3 \leftrightarrow \alpha_2)}(B')+E_{\beta_4 \leftrightarrow \alpha_3}(B') = 2 (-g_J \mu_B + g_d \mu_K) B'$). Thus, the sum of the hyperfine splitting energies allows one a precise measurement of the g-factors and to test the actual QED corrections~\cite{moskovkin} of the Breit-Rabi formula that are on a level of $10^{-3}$ up to $10^{-2}$ of the values for the observable transitions at these small magnetic fields below 5 mT.\\
This new type of spectroscopy is a beautiful demonstration to describe experimental results with the tools of quantum mechanics and allows one to determine and to observe the smallest direct transitions between two quantum states. With lower beam velocities even smaller energy differences are possible until the quality of the magnetic field configuration is sufficient. This precision opens the door not only to QED tests of the hyperfine splitting energies but even to a new type of parity-violation experiments with hydrogen and deuterium atoms \cite{Heidelberg}.
\newpage
\begin{acknowledgments}
The authors wish to thank Alexander Milstein, Yurij Shestakov and Dimitri Toporkov from the Budker Institute in Novosibirsk and Alexander Belov from the Russian Academy of Science in Moscow for valuable discussions.\\
The work has received funding from the ATHENA project of the Helmholtz Association (HGF). Lukas Huxold acknowledges support by the Deutsche Forschungsgemeinschaft within project BU 2227/1-1.
\end{acknowledgments}

\end{document}